\documentclass[12pt,english,floatfix,superscriptaddress,aps,preprint,amsmath,amssymb]{revtex4}
\usepackage[latin9]{inputenc}
\usepackage{amsmath}
\usepackage{amssymb} % For \hslash
\usepackage{amsbsy}
\usepackage{amsfonts}
\usepackage{amsopn}
\usepackage{amstext}
\usepackage{graphicx}
\usepackage{esint}
\usepackage{hyperref}

\usepackage{amssymb}
\usepackage{amsfonts}
\usepackage{amsmath}
\usepackage{graphicx}
\usepackage[english]{babel}
\usepackage{color}
\usepackage[dvips]{epsfig}
\usepackage[dvips]{graphicx}
\usepackage{float}
\usepackage{units}
\usepackage{textcomp}
\usepackage{babel}

\def \be {\begin{equation}}
\def \ee {\end{equation}}
\def \bea{\begin{eqnarray}}
\def \eea{\end{eqnarray}}
\def \ba {\begin{align}}
\def \ea {\end{align}}

\begin{document}

\title{Dark Energy and Normalization of the Cosmological Wave Function}
\author{Peng Huang}
\email{huangp46@mail.sysu.edu.cn}
\affiliation{School of Astronomy and Space Science, Sun Yat-Sen University, Guangzhou 510275, People's Republic of China}

\author{Yue Huang}
\email{huangyue@itp.ac.cn}
\affiliation{State Key Laboratory of Theoretical Physics, Institute of Theoretical Physics, Chinese Academy of Sciences, Beijing, 100190}
\affiliation{Kavli Institute for Theoretical Physics China, Chinese Academy of Sciences, Beijing, 100190}

\author{Miao Li}
\email{limiao9@mail.sysu.edu.cn}
\affiliation{School of Astronomy and Space Science, Sun Yat-Sen University, Guangzhou 510275, People's Republic of China}
\affiliation{State Key Laboratory of Theoretical Physics, Institute of Theoretical Physics, Chinese Academy of Sciences, Beijing, 100190}

\author{Nan Li}
\email{linan@itp.ac.cn}
\affiliation{State Key Laboratory of Theoretical Physics, Institute of Theoretical Physics, Chinese Academy of Sciences, Beijing, 100190}
\affiliation{Kavli Institute for Theoretical Physics China, Chinese Academy of Sciences, Beijing, 100190}

\begin{abstract}
Dark energy is investigated from the perspective of quantum cosmology. It is found that, together with an appropriate normal ordering factor $q$, only when there is dark energy then can the cosmological wave function be normalized. This interesting observation may require further attention.
%By treating the existence of a classical universe as a constraint, it is found that the normal ordering ambiguity factor $q$ in Wheeler-DeWitt equation tends to take its value on domain $(-1, 3)$. Furthermore, to ensure the existence of a classical universe, there must be dark energy in the universe. It is  in this sense we propose that dark energy is the reason for the existence of a classical universe.
\end{abstract}
\maketitle

The accelerating expansion of our universe has presented a very challenging problem in theoretical physics as well as in cosmology \cite{Weinberg:1988cp,Zlatev:1998tr}.
We believe that the character of dark energy is deeply associated with the nature of quantum gravity. In fact, one of the authors showed that  dark energy can be well described with the help of the holographic principle, a characteristic feature of any viable theory of quantum gravity. This holographic dark energy model \cite{Li:2004rb} becomes one of the most competitive and popular dark energy candidates, which also implies that a deep relation between dark energy and quantum gravity is well appreciated in the community.

On the other hand, as an application of quantum physics to the dynamical systems describing closed universes, quantum cosmology is also tightly related to quantum gravity. In spite of some misgivings, interesting hints on fundamental mathematical and physical questions can be obtained from quantum cosmology. The studies on quantum cosmology as a generally covariant and highly interacting quantum theory may also provide answers to questions concerning the interplay of symmetries, discrete structures and so forth.

Having noticed the common connection between dark energy, quantum cosmology and quantum gravity, a natural and interesting question to ask is whether there is relation between dark energy and quantum cosmology. An answer to this problem may deepen our understanding of the dark energy and shed light on problems related to it.\\
\\
For a theory which still retains its highly speculative and controversial nature, it is necessary to clarify the notation we will use in the following at first. For a general Wheeler-DeWitt equation for quantum cosmology
\begin{equation}
\label{1}
[-\hbar^2 \frac{\partial^{2}}{\partial a^{2}}-\hbar^2 \frac{q}{a}\frac{\partial
}{\partial a}+\frac{9\pi^{2}}{4G^{2}}(  a^{2}-\frac{\Lambda}{3}
a^{4})+\mathcal{H}_{matter}]\Psi(a)=0,
\end{equation}
where we assume a positive spatial curvature as refelcted in the term proportional to $a^2$. We define the inner product of the wave function as
\begin{equation}
\label{2}
\mathcal{P}=\int {a^q|\Psi|^2da}
\end{equation}
and interpret $d\mathcal{P}$ as the probability that the universe stays within scale factor between $a$ and $a+da$ when the wave function is normalizable.
This quantity will be of our central interest.  Of course, the probability interpretation is not simple, for instance to return to usual quantum mechanics
one need to introduce time, as proposed in \cite{Weinberg:1988cp}, one can imagine introducing  a scalar field to play the role of time.

To focus on the key problem without lose of generality, it is convenient to suppose that the universe is closed and filled only with a special dark energy which is just the cosmological constant. In this setup, the Wheeler-DeWitt equation has a simpler form as
\begin{equation}
\label{3}
[-\frac{\partial^{2}}{\partial a^{2}}-\frac{q}{a}\frac{\partial
}{\partial a}+\frac{1}{l_p^4}(  a^{2}-\frac{a^{4}}{l_{\Lambda}^2}
)]\Psi(a)=0
\end{equation}
with $l_p=(\frac{4G^{2} \hbar^2}{9\pi^{2}})^{\frac{1}{4}}$, $l_{\Lambda}=(\frac{3}{\Lambda})^{\frac{1}{2}}$ corresponding to Planck scale and length scale introduced by the cosmological constant, respectively.

In the region $a\ll l_{\Lambda}$, the energy induced by spatial curvature dominates the universe, thus the term $\frac{a^{4}}{l_{\Lambda}^2}
$ in Eq.(\ref{3}) can be neglected. In this situation, the general solution of Eq.(\ref{3}) has the form as
\begin{equation}
\renewcommand{\arraystretch}{1.5}
\label{sol}
\left\{
\begin{array}{ll}
\Psi(a)=C_1(\frac{a}{l_p})^{\frac{1-q}{2}}I_{\frac{q-1}{4}}(\frac{a^2}{2l_p^2})+C_2(\frac{a}{l_p})^{\frac{1-q}{2}}K_{\frac{q-1}{4}}(\frac{a^2}{2l_p^2}), \quad for \quad q\geq 1;\\
\Psi(a)=C_1(\frac{a}{l_p})^{\frac{1-q}{2}}I_{\frac{1-q}{4}}(\frac{a^2}{2l_p^2})+C_2(\frac{a}{l_p})^{\frac{1-q}{2}}K_{\frac{1-q}{4}}(\frac{a^2}{2l_p^2}), \quad for \quad  q<1.
\end{array}
\right.
\end{equation}
It is well known that, for $0<\mid z\mid\ll\sqrt{\alpha+1}$, the asymptotic behavior for modified Bessel functions is
\begin{equation}
I_{\alpha}(z)\sim\frac{1}{\Gamma(\alpha+1)}(\frac{z}{2})^{\alpha};
\end{equation}
\begin{equation}
K_{\alpha}(z)\sim
\left\{
\begin{array}{ll}
-\ln (\frac{z}{2})-\gamma, \quad for \quad  \alpha=0, \nonumber \\
\frac{\Gamma(\alpha)}{2}(\frac{2}{z})^{\alpha}, \qquad for \quad \alpha>0.
\end{array}
\right.
\end{equation}
These results imply three following asymptotic behaviors for modified Bessel functions which are of our concern:
\begin{equation}
I_{\frac{q-1}{4}}(\frac{1}{2} (\frac{a}{l_p})^2) \sim (\frac{a}{l_p})^{\frac{q-1}{2}}, \quad K_{\frac{q-1}{2}}(\frac{1}{4} (\frac{a}{l_p})^2) \sim (\frac{a}{l_p})^{\frac{1-q}{2}}, \quad for \quad q>1;
\end{equation}
\begin{equation}
I_{0}(\frac{1}{2} (\frac{a}{l_p})^2) \sim A_1, \quad K_{0}(\frac{1}{2} (\frac{a}{l_p})^2) \sim \ln{a}+A_2, \quad for \quad q=1;
\end{equation}
\begin{equation}
I_{\frac{1-q}{4}}(\frac{1}{2} (\frac{a}{l_p})^2) \sim (\frac{a}{l_p})^{\frac{1-q}{2}}, \quad K_{\frac{1-q}{2}}(\frac{1}{4} (\frac{a}{l_p})^2) \sim (\frac{a}{l_p})^{\frac{q-1}{2}},\quad for \quad q<1.
\end{equation}
After some tediously but straightforward calculations, one can find that the second term of the wave function (\ref{sol}) will always make the inner product $\mathcal{P}=\int^{\epsilon}_0 {a^q|\Psi|^2da}$ ($\epsilon$ is an arbitrary upper limit of the integral satisfying $\epsilon \ll l_{\Lambda}$) divergent unless the normal ordering ambiguity factor $q$ is in the domain of $(-1,3)$.

Another way to investigate the Wheeler-DeWitt equation in the region $a\ll l_{\Lambda}$ is to use WKB method. Since the Wheeler-DeWitt equation is a second order differential function, boundary conditions are used in this method to select a particular solution. Two most famous wave functions are Hartle-Hawking  no-boundary wave function \cite{Hartle:1983ai} and Vilenkin's tunnelling wave function \cite{Vilenkin:1984wp,Farhi:1988qp}. It will be interesting to find out to which solutions their proposals correspond in the general solution space.

To proceed, one can write
$\Psi(a)$ as $\Psi(a)=e^{i \frac{S}{l_p^2}}$ with $S=S_0+l_p^2 \; S_1+\mathcal{O}(l_p^4)$. Inserting this into Eq.(\ref{3}) and arranging terms according to order in $l_p$, one obtains the zeroth order and second order equations,
\begin{equation}
\partial_a S_0=\pm i \sqrt{a^2-\frac{a^4}{l_{\Lambda}^2}}, \quad for \quad a<l_{\Lambda};
\end{equation}
\begin{equation}
\partial_a S_0 \partial_a S_1=\frac{i}{2}(\partial_a^2 S_0+\frac{q}{a}\partial_a S_0).
\end{equation}
The  above two equations can be solved to obtain the WKB wave function
\begin{equation}
\Psi(a)_{\pm}=e^{i (\frac{S_0}{l_p^2}+S_1)}=a^{-\frac{q}{2}} \frac{1}{\sqrt{p}} e^{\pm \frac{1}{l_p^2}\int_{a}^{l_{\Lambda}} da' p(a')},\quad for \quad a<l_{\Lambda},
\end{equation}
with $p(a)=\sqrt{a^2-\frac{a^4}{l_{\Lambda}^2}}$. We have denoted the two linearly-independent solutions as ``$\pm$" according to Vilenkin. It is then easy to see that Hartle-Hawking's choice was $\Psi_-$ while Vilenkin's choice was $\Psi_+-\frac{i}{2} \Psi_-$ \cite{Vilenkin:1984wp}. However, Vilenkin argued that $\Psi_-$ is negligible except in the vicinity of $l_{\Lambda}$ so that he actually used $\Psi_+$ in the region $a<l_{\Lambda}$. Notice that the behavior of wave function as $a\rightarrow0$ is in fact not derivable from the WKB wave function because the WKB method is valid in the region $a\gg l_p$. Thus, to see the true behavior of wave functions according to their choices, one needs to figure out what wave functions they chose out of the general solution (\ref{sol}). The strategy is to match Hartle-Hawking's and Vilenkin's WKB wave functions with (\ref{sol}) in a region where both the WKB solution and asymptotic solution are valid, i.e. the region $l_p \ll a \ll l_{\Lambda}$. In this region, the WKB wave functions can be approximated by
\begin{equation}
\Psi_{\pm} \approx a^{-\frac{q+1}{2}} e^{\pm \frac{1}{3} (\frac{l_{\Lambda}}{l_p})^2} e^{\mp \frac{1}{2}(\frac{a}{l_p})^2}
\end{equation}
For the asymptotic solution (\ref{sol}), one needs to use the asymptotic form of modified Bessel functions for large arguments
\begin{equation}
I_{\alpha}\sim \frac{e^z}{\sqrt{2 \pi z}}(1-\frac{4 z^2-1}{8 z}+\mathcal{O}(z^{-2})), \quad for \quad |argz|<\frac{\pi}{2},
\end{equation}
\begin{equation}
K_{\alpha}\sim \frac{e^{-z}}{\sqrt{2 \pi z}}(1+\frac{4 z^2-1}{8 z}+\mathcal{O}(z^{-2})), \quad for \quad  |argz|<\frac{3 \pi}{2}.
\end{equation}
So Eq.(\ref{sol}) approximates to
\begin{equation}
\Psi \sim C_1 a^{-\frac{q+1}{2}} e^{\frac{1}{2}(\frac{a}{l_p})^2}+C_2 a^{-\frac{q+1}{2}} e^{-\frac{1}{2}(\frac{a}{l_p})^2}
\end{equation}
Now it can be seen that Hartle-Hawking's choice corresponds to the modified Bessel function of second kind $I$ while Vilenkin's choice corresponds to the modified Bessel function of first kind $K$. It could be seen that Hartle-Hawking's wave function leads to a convergent probability whatever $q$ is, while the inner product of Vilenkin's wave function is convergent only when $-1<q<3$.\\
\\
In the region $a\gg l_{\Lambda}$, the cosmological constant term prevails over spatial curvature, the term $a^2$ in Eq.(\ref{3}) can be neglected. The general solution of Eq.(\ref{3}) then is
\begin{equation}
\label{t1}
\Psi(a)=C_1 (\frac{a^3}{l_p^2 l_{\Lambda}})^{\frac{1-q}{6}} J_{\frac{q-1}{6}}(\frac{1}{3} \frac{a^3}{l_p^2 l_{\Lambda}})+C_2 (\frac{a^3}{l_p^2 l_{\Lambda}})^{\frac{1-q}{6}} Y_{\frac{q-1}{6}}(\frac{1}{3} \frac{a^3}{l_p^2 l_{\Lambda}}).
\end{equation}
Using the asymptotic expansion of Bessel functions at large argument as
\begin{equation}
J_{\alpha}(z)=\sqrt{\frac{2}{\pi z}}[\cos(z-\frac{\alpha \pi}{2}-\frac{\pi}{4})+e^{|Im(z)|}\mathcal{O}(|z|^{-1})], \quad for \quad |argz|<\pi,
\end{equation}
\begin{equation}
Y_{\alpha}(z)=\sqrt{\frac{2}{\pi z}}[\sin(z-\frac{\alpha \pi}{2}-\frac{\pi}{4})+e^{|Im(z)|}\mathcal{O}(|z|^{-1})], \quad for \quad |argz|<\pi,
\end{equation}
one can get
\begin{equation}
\label{t3}
J_{\frac{q-1}{6}}(\frac{1}{3}\frac{a^3}{l_p^2 l_{\Lambda}}) \sim a^{-\frac{3}{2}} \cos(\frac{1}{3}\frac{a^3}{l_p^2 l_{\Lambda}}+phase);
\end{equation}
\begin{equation}
\label{t4}
Y_{\frac{q-1}{6}}(\frac{1}{3}\frac{a^3}{l_p^2 l_{\Lambda}}) \sim a^{-\frac{3}{2}} \sin(\frac{1}{3}\frac{a^3}{l_p^2 l_{\Lambda}}+phase).
\end{equation}
Then it is easy to find that the inner product $\mathcal{P}=\int^{+\infty}_{\lambda} {a^q|\Psi|^2da}$ ($\lambda$ is an arbitrary lower limit of the integral satisfying $\lambda \gg l_{\Lambda}$) always converges regardless of the normal ordering factor ambiguity $q$.

The interesting thing is, together with an appropriate normal ordering factor $q$, cosmological-constant dark energy causes a convergent and thus normalizable cosmological wave function. It is natural to wonder whether the relation between cosmological-constant dark energy and normalization of cosmological wave function is only a coincidence or a profound observation which deserves more attention. To investigate this,
%Till now, the importance of the value of the normal ordering factor is demonstrated. Nevertheless, the significance of the cosmological-constant dark energy for the existence of the classical universe is not explicit since all it does is to cause a convergent probability for the classical universe. It is quite possible that other energy components will also lead to a probability which can ensure the quantum universe grows into classical one. To investigate such possibilities in detail,
we consider following Wheeler-DeWitt equation
\begin{equation}
\label{6}
[-\frac{\partial^{2}}{\partial a^{2}}-\frac{q}{a}\frac{\partial
}{\partial a}+\frac{1}{l_p^4}( a^{2}-\frac{1}{f(h)} \frac{a^h}{l_p^{h-2}})]\Psi(a)=0
\end{equation}
with $f(h)$ a general function of parameter $h$ and $f(4)=(\frac{l_{\Lambda}}{l_p})^2$ is required. Apparently, term $\frac{1}{f(h)} \frac{a^h}{l_p^{h-2}}$ corresponds to a general energy component whose index of equation of state has the value of $w=\frac{1-h}{3}$. Thus, $h=0, 1, 4$ corresponds to radiation, matter and cosmological constant, respectively.
When $h>2$, in the region $a\rightarrow 0$, it is easy to find that the energy induced by spatial curvature again dominates the universe, thus the results got here will be the same as that for Eq.(\ref{3}), i.e., the inner product for the wave function corresponding to the universe described by Eq.(\ref{6}) with small scale factor is convergent only when the normal ordering ambiguity factor $q$ takes its value on the domain of $(-1,3)$. For large $a$, Eq.(\ref{6}) becomes
\begin{equation}
[-\frac{\partial^{2}}{\partial a^{2}}-\frac{q}{a}\frac{\partial
}{\partial a}-\frac{1}{f(h)} \frac{a^h}{l_p^{h+2}})]  \Psi(a)  =0.
\end{equation}
Its general solution is
\begin{equation}
\Psi(a)=C_1 a^{\frac{1-q}{2}} J_{\frac{q-1}{h+2}} (\frac{2}{h+2}\; \frac{a^{\frac{h+2}{2}}}{l_p^{\frac{h+2}{2}} \sqrt{f}})+C_2 a^{\frac{1-q}{2}} Y_{\frac{q-1}{h+2}} (\frac{2}{h+2}\; \frac{a^{\frac{h+2}{2}}}{l_p^{\frac{h+2}{2}} \sqrt{f}}).
\end{equation}
Using the asymptotic forms of Bessel functions given in Eq.(\ref{t3}) and Eq.(\ref{t4}), one can find that the asymptotic form of the general solution is
\begin{equation}
\Psi(a)\sim C_1 a^{-\frac{h+2}{4}} \cos[\frac{2}{h+2} \frac{a^{\frac{h+2}{2}}}{l_p^{\frac{h+2}{2}}\sqrt{f}}+phase]+C_2 a^{-\frac{h+2}{4}} \sin[\frac{2}{h+2} \frac{a^{\frac{h+2}{2}}}{l_p^{\frac{h+2}{2}}\sqrt{f}}+phase].
\end{equation}
Then it can be shown that the integral of probability still converges on the large $a$ side for $h>2$. It can also be easily found that, after a similar process, the wave function is divergent when $h=2$. When $h<2$, the asymptotic wave function on the large $a$ side becomes the same as Eq.(\ref{sol}). Notice that for large argument, the modified Bessel function of the first kind has an exponentially divergent part while the modified Bessel function of the second kind has a exponentially convergent part. So the integral of generic probability for $h<2$ is divergent.

What has been shown is that, for an energy component whose index of equation of state is $w=\frac{1-h}{3}$, the
%inner product of the wave function is convergent
cosmological wave function is normalizable only if the energy component is dark energy (with $h>2$).
%The interesting thing is, a normalizable wave function is a precondition for us to observe a classical universe in \textit{finite} region. To see this, let us first consider a particle on an infinite plane. For a free particle with definite momentum, its wave function is of course the plane wave, which is not normalizable. The non-normalizability of this wave function tells us that the probability of finding the particle in a finite range is actually zero. If we instead consider a particle in a finite box, then its wave function will be a normalizable one, which gives us a non-vanishing probability of finding this particle in a finite range. Similarly, one can claim the existence of the classical universe only when one has observed a classical universe in \textit{finite} region. However, if the wave function for this classical universe is nonnormalizable, the probability to find a classical universe in \textit{finite} region is zero. Thus, the wave function corresponding to the classical universe must be normalizable. The key point is, only a dark energy can lead the inner product of the cosmological wave function to be convergent. On the contrary, ordinary energy components, such as matter or radiation, will always lead to a divergent inner product. Thus, dark energy is essential to ensure an observable classical universe, in this sense, one can say that  dark energy is the reason for the existence of the classical universe.\\
It needs to be stressed that the above claim holds well only when the requirement $w<-\frac{1}{3}$ is always satisfied when $a$ is large enough. To see this, let us consider an general energy component which leads to a Wheeler-DeWitt equation as following
\begin{equation}
\label{t5}
[-\frac{\partial^{2}}{\partial a^{2}}-\frac{q}{a}\frac{\partial
}{\partial a}+\frac{1}{l_p^4}(a^{2}-l_p^2g(\frac{a}{l_p}))]\Psi(a)=0
\end{equation}
with $g(\frac{a}{l_p})$ an function of $\frac{a}{l_p}$ which will prevail over $a^2$ when $a$ is large enough. Denoting $\Psi(a)=e^{i(\frac{S_0}{l^2_p}+S_1)}$, and supposing that the condition $|S'|^2\gg l_p^2|S''+\frac{q}{x}S'|$, which takes the following form for our case
\begin{equation}
\label{t9}
g-x^2\gg\frac{1}{2}\frac{g'-2x}{g-x^2}+\frac{q}{x},
\end{equation}
is satisfied by $g(x$), then through a general procedure of the WKB method, one get the wave function and the inner product corresponding to it as following
\begin{equation}
\Psi(a)=(S_0^{'}x^2)^{-\frac{1}{2}}e^{i\int{\sqrt {g-x^2}da}},
\end{equation}
\begin{equation}
\label{t7}
\mathcal{P}\propto \int {\frac{dx}{\sqrt{g-x^2}}}
\end{equation}
with $x=\frac{a}{l_p}$ and the prime denoting differentiation with respect to $x$. Since $g(\frac{a}{l_p})$ prevails over $a^2$ when $a$ is large enough, Eq.(\ref{t7}) can be simplified further into
\begin{equation}
\Psi(a)=(S_0^{'}x^2)^{-\frac{1}{2}}e^{i\int{\sqrt {g}da}},
\end{equation}
\begin{equation}
\label{t8}
\mathcal{P}\propto \int {\frac{dx}{\sqrt{g}}}.
\end{equation}

Now, considering a special energy component whose appearance in Wheeler-DeWitt equation is $g(x)=x^2\ln x$, the special character of this kind of energy component is that its index of equation of state is $w=-\frac{1}{3}-\frac{1}{3\ln x}$ which will always be less than $-\frac{1}{3}$ (for $a$ is large) except at the point where the scale factor $a$ is infinite. One can check that the term $g(x)=x^2\ln x$ indeed prevails over the term $a^2$ in Eq.(\ref{t5}) and it also satisfies condition (\ref{t9}), thus, one can safely insert it into (\ref{t8}) to find that the inner product is divergent. Now, considering another energy component for which the condition $g(x)>x^{2+\epsilon}$ for a large scale factor ($\epsilon$ is an arbitrary given positive constant) is satisfied. This function corresponds to an energy component with index of equation of state $w=-\frac{1+\epsilon}{3}$ which is less than $-\frac{1}{3}$ when $a$ is large enough. Then, after a similar process, one can find
\begin{equation}
\mathcal{P}\propto\int_{\lambda}^{+\infty} {\frac{dx}{\sqrt{g}}}<\int_{\lambda}^{+\infty} {\frac{dx}{x^{\frac{\epsilon}{2}+1}}}=\frac{2}{\epsilon}\lambda^{-\frac{\epsilon}{2}},
\end{equation}
which is convergent.

The difference between the first and the second energy component considered above is that, while the second energy component always satisfies the condition $w<-\frac{1}{3}$ for large scale factor, the first energy component violates it at point $a=\infty$. Then, despite the first energy component behaves precisely like ordinary dark energy most of the time in the sense that it can accelerate the universe, it still leads to a divergent inner product of the wave function because it deviates from dark energy at the point with $a=\infty$. Thus, as having been stressed, the claim that dark energy causes a normalizable cosmological wave function only holds in the circumstances that $w<-\frac{1}{3}$ is strictly satisfied when $a$ is large enough.\\
\\
\qquad To conclude, we study dark energy from the perspective of quantum cosmology in this work. It is found that, together with an appropriate normal ordering factor $q$ (take value on domain $(-1, 3)$, dark energy leads to a normalizable cosmological wave function while other energy components can not. We think that the interesting relation between dark energy and normalization of cosmological wave function may imply some deep characters of dark energy which deserves more attention. For further investigation, it is an interesting question whether or not quantum cosmology gives us some hints on choosing the dark energy models. There might not be a good answer based on the cases considered in this paper as the functions representing dark energy are rather arbitrary. However, we believe that there exists an interesting answer to this question when the dynamics of the dark energy is known (in our case, when the dynamics of $h$ is known). Furthermore, considering the fact that the exact meaning of the wave function of the universe and the problem of measurement is still under careful investigation in quantum cosmology, it may be interesting to consider this work from the light of a more precise version of quantum theory, like many worlds or de Broglie-Bohm theory.

\begin{acknowledgments}
We thank the anonymous referee for the valuable and insightful communication. ML is supported by the National Natural Science Foundation of China (Grant No. 11275247, and Grant No. 11335012) and 985 grant at Sun Yat-Sen University (Grant No. 99122-18811301). YH and NL would like to thank the support of the School of Astronomy and Space Science, Sun Yat-Sen University.
\end{acknowledgments}

\end{document}